\documentclass[pre,floatfix,twocolumn,nopacs]{revtex4}
\usepackage{amssymb}
\usepackage{amsmath}
\usepackage{graphicx}
\usepackage{color}
\usepackage{float}

\begin{document}

\title{Influence of different disorder types on Aharonov-Bohm caging in the diamond chain}
\author{Goran Gligori\'c$^{1}$, Daniel Leykam$^{2}$, Aleksandra Maluckov$^{1}$}
\affiliation{$^{1}$P$^{*}$ Group, Vin\v ca Institute of Nuclear Sciences, University of Belgrade, P.O. Box 522, 11001 Belgrade, Serbia\\
$^{2}$Center for Theoretical Physics of Complex Systems, Institute for Basic Science, Daejeon 34126, Republic of Korea}

\begin{abstract}
The linear diamond chain with fine-tuned effective magnetic flux has a completely flat energy spectrum and compactly-localized eigenmodes, forming an Aharonov-Bohm cage. We study numerically how this localization is affected by different types of disorder (static and time-evolving) relevant to recent realizations of Aharonov-Bohm cages in periodically-modulated optical waveguide arrays. We demonstrate robustness of localization under static and periodically-evolving disorder, while in contrast non-quenched (time-dependent) disorder leads to wavepacket spreading and delocalization. 

\end{abstract}

\maketitle
\section{Introduction}

The fundamental phenomenon of wave localization can arise from a variety of sources. In nonlinear media, modulational instability leads to localization in the form of wave self-trapping and solitons \cite{dd}, \cite{dd1}. In linear media, time-independent quenched disorder (QD) induces localization via interference effects in a process known as Anderson localization~\cite{Anderson1958}. By contrast, randomly time-evolving non-quenched disorder (NQD) can result in delocalization and enhanced wave transport~\cite{SegevHypertransport,SegevSuperdiffusion}. Linear localization even without disorder occurs in certain periodic structures known as flatband (FB) networks \cite{p21,quantrings,moller,fang}. In the FB networks perfectly localized compact modes arise due to the lattice geometry enabling destructive interference between different propagation paths~\cite{FB_review,FB_review2}.  One of the most intriguing questions is to understand the interplay between these different localization mechanisms~\cite{FlachNL,nasabc,diliberto}. For example, the combination of nonlinearity and disorder induces delocalization and subdiffusive wave spreading via nonlinear interactions between Anderson-localized linear modes~\cite{FlachNL}. 

In the case of FB networks QD transforms the compact modes into Anderson-localized modes with exponential tails~\cite{EPLFlach}. Statistical properties of the localized modes are sensitive to the location of the FBs within the lattice Bloch wave spectrum. For example, when FBs coexist at the same energy as other non-flat (dispersive) bands, Fano resonance-like interaction between the bands leads to ``heavy-tailed'' statistics in which the mode localization is chiefly determined by rare realizations of the disorder~\cite{Flach} and the FB states acquire a finite lifetime~\cite{Gneiting2018}. On the other hand, spectrally-isolated FBs typically exhibit strong localization insensitive to the precise strength or profile of the QD, whether on-site (local disorder) describing randomness in the site energies or inter-site (hopping) disorder~\cite{Flach}. 

In this article we study the influence of various forms of disorder on localization in a particular FB network: the one-dimensional Aharonov-Bohm (AB) cage~\cite{abohm,vidal}. AB cages are peculiar structures in which all of the spectral bands are perfectly flat. Originally generated by applying a strong magnetic field to superconducting wire networks~\cite{vidal2}, more recently AB cages were realized as photonic waveguide arrays by generating an effective magnetic flux using periodic modulation of the waveguides along the propagation axis~\cite{ahr,modulatedPL, QDSzamait,Longinonherm,Mukherjee,Szameit}. This longitudinal modulation is analogous to time-dependent modulation of quantum particles described by the Schr\"odinger equation, and therefore we consider both static and time-dependent sources of disorder. Disorder can evolve periodically (forming a variant of QD) or non-periodically in time (standard meaning of the NQD). The compact mode behaviour in such cases has not been explored before. These two cases are also intriguing  because the mathematical interpretation of localization requires consideration of the time-dependent Hamiltonians, which is usually not the case with respect to the Anderson localization. The main question we are interested in is whether the time-dependent disorder can lead to delocalization, as is the case for regular (non-FB) lattices~\cite{SegevHypertransport,SegevSuperdiffusion}. 

We model periodically-evolving disorder by assuming an initial QD profile has a sinusoidal time dependence [Eq.~\eqref{disorder}]. To obtain non-periodic disorder we randomly change the phase of this sinusoidal modulation at regular time intervals. We find numerically that periodic QD leads to similar strong localization to the standard QD, which can be understood in terms of localized Floquet eigenstates. On the other hand, non-periodic disorder leads to delocalization, even in this extreme limit where all  bands are perfectly flat. 

The outline of the paper is as follows: In Sec.~\ref{sec:model} we make a brief overview of the studied phenomena and establish corresponding mathematical model and the methods of numerical analysis. We then study the time evolution of a compact mode excitation in the presence of different disorder types: Sec.~\ref{sec:static} reviews the case of static disorder, which does not disrupt the strong localization of the eigenstates. We find the chain is more sensitive to hopping disorder than the on-site disorder. Sec.~\ref{sec:periodic} analyzes the case of periodically evolving disorder, and Section~\ref{sec:nqd} discusses the NQD. The concluding Sec.~\ref{sec:conclusion} summarizes our main findings. 

\section{Model and methods}
\label{sec:model}

\begin{figure}

\includegraphics[width=0.75\columnwidth]{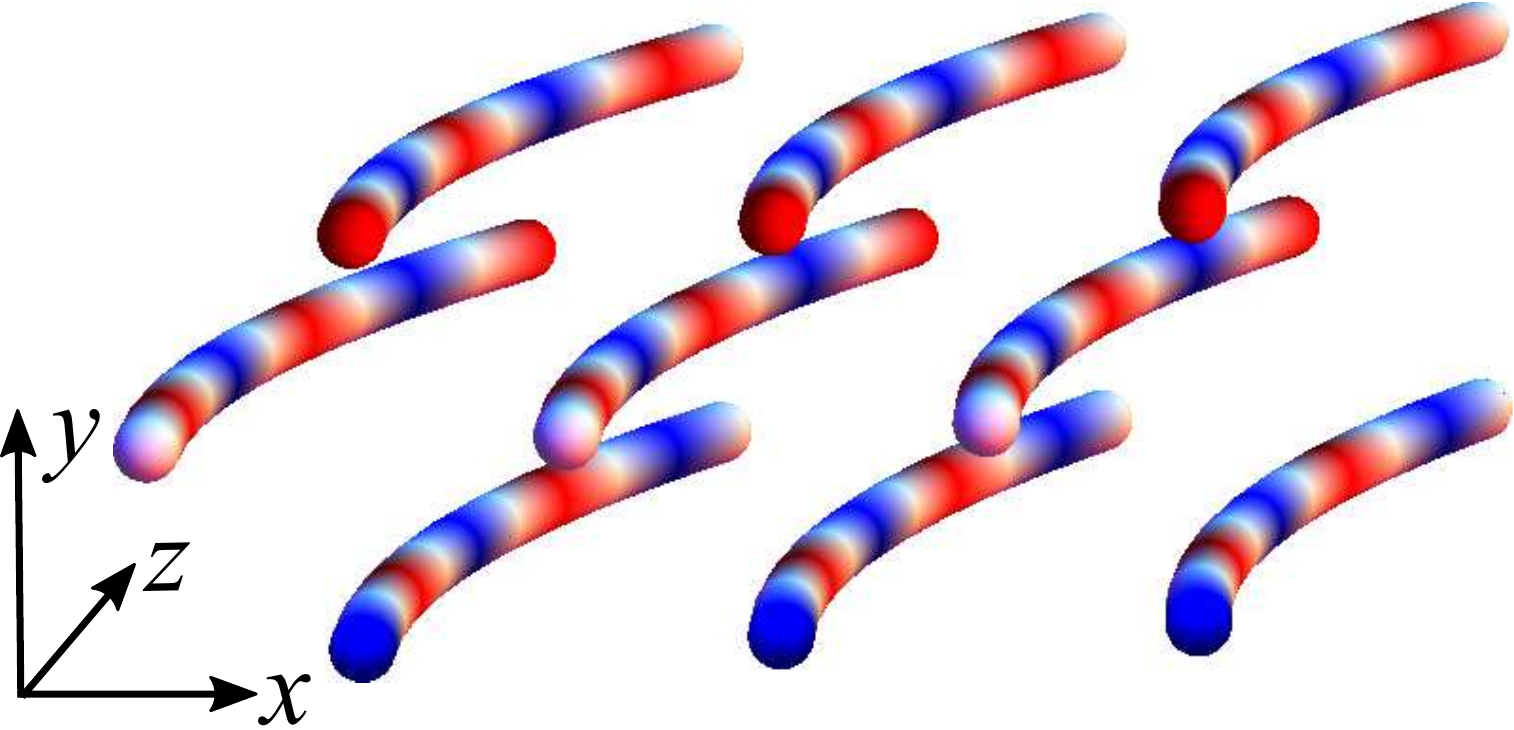}

\caption{Schematic of optical waveguide array with effective magnetic flux implemented via periodic modulation of the waveguide depths (shown in red and blue) combined with transverse acceleration of the waveguide positions.}

\label{fig:schematic}

\end{figure}

AB caging was first studied in Ref.~\cite{vidal} for tight-binding
electrons in certain two-dimensional lattices and quasi-one-dimensional lattices, including the diamond chain. The perfectly flat spectrum arises from destructive interference
among different hopping paths at a critical value of the
magnetic flux. AB caging was experimentally demonstrated in superconducting wire networks~\cite{vidal2}, mesoscopic
semiconductor lattices \cite{26}, and arrays of Josephson
junctions \cite{27}. Here we consider
light transport in a periodic diamond chain of coupled optical waveguides in the presence of an artificial gauge field induced by periodically modulating the waveguides, illustrated schematically in Fig.~\ref{fig:schematic}.

Following the procedure presented in Ref.~\cite{ahr}, we model the evolution of slowly varying optical field amplitudes in the waveguides, $(a_n,b_n,c_n)$ as
\begin{eqnarray}
i\frac{da_n}{dz}&=&\kappa\left( b_n\,e^{-i\Gamma_{bn}/2} +
b_{n-1}
+c_n+c_{n-1}\,e^{-i\Gamma_{cn}/2}\right) + \nonumber\\ & & +\epsilon_{an}a_n, \nonumber\\
i\frac{db_n}{dz}&=&\kappa\left( a_n\,e^{i\Gamma_{bn}/2} +
a_{n+1}\right)+\epsilon_{bn}b_n, \label{system6} \\ 
i\frac{dc_n}{dz}&=&\kappa\left( a_n +
a_{n+1}\,e^{i\Gamma_{cn+1}/2} \right)+\epsilon_{cn}c_n,\nonumber 
\end{eqnarray}
where $n=1,.., N$ indexes the cells, $\Gamma_n$ is the effective magnetic flux in
each plaquette, and $\kappa$ is the coupling coefficient
\cite{ahr,nasabc}, fixed to $1$ in the numerical simulations without loss of generality. These equations are obtained by averaging over high frequency modulation in $z$ and can thus host both QDs and NQDs.

The $\epsilon_{jn}$ terms describe the on-site disorder potential, which we model as~\cite{nashkagome}
\begin{equation}
\epsilon_{jn}=\epsilon_{jn}^{QD}+\epsilon_{jn}^{ED}(z)=\epsilon_{jn}^{(0)}\left(1+A\sin\left(\omega_{0}z+\phi_{jn}(z)\right)\right),\label{disorder}
\end{equation}
where $j=a,b,c$ is the sublattice index. The term $\epsilon_{jn}^{QD} \equiv \epsilon_{jn}^{(0)}$ describes the on-site static disorder, while $\epsilon_{jn}^{ED}(z)$ describes the on-site evolving
disorder. We take $\epsilon_{jn}^{(0)}$ to be uncorrelated random numbers from the interval
$\left[-W/2,W/2\right]$, where the parameter $W$ is the disorder strength.  The time evolving disorder term is characterized by amplitude $A$, frequency $\omega_{0}=2\pi/Z_{0}$, and the phase
terms $\phi_{jn}(z)$.

Meanwhile, we model hopping disorder as fluctuations in the hopping phase determining the effective magnetic flux, 
\begin{equation}
\Gamma_{jn}=\Gamma_0\left(1+\delta \Gamma_{jn}(z)\right).
\end{equation}
The last term $\delta \Gamma_{jn}(z)$ takes a similar form to $\epsilon_{jn}(z)$, being separated into static and $z$-dependent terms. The small fluctuations of the magnetic
flux resemble the experimental uncertainties in preparing the
external artificial magnetic field. 

We further distinguish the evolving part of the disorder into two different classes: whether it evolves periodically or non-periodically in $z$. The latter models NQD, which can be interpreted as an uncorrelated thermally-induced disorder in the context of electronic systems. To create NQD with controllable rate of change in $z$, we follow the approach of Ref.~\cite{teza} and consider noise in the modulation phase $\phi_{jn}(z)$. After a fixed step length, named the dephasing length $\Delta z$, we randomly reassign the modulation phases $\phi_{jn}(z)$, drawn uniformly from the interval $[-\pi,\pi]$. 

\begin{figure}[h]
    \center\includegraphics [width=\columnwidth]{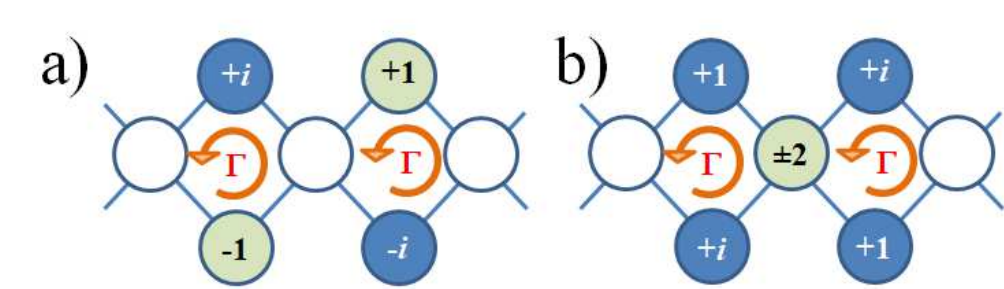} \caption{
        Schematic representation of FB modes in the diamond chain. A
        synthetic magnetic flux $\Gamma$ is applied in each plaquette. Geometry-induced $\beta_{FB} = 0$ FB mode (a)  and  additional FB modes 
        ($\beta_{FB}=\pm 2$) (b) originating due to the AB caging.}
    \label{fig1}
\end{figure}

The eigenvalue spectrum of the diamond chain with artificial flux $\Gamma$ in the absence of disorder consists
of three bands:
\begin{equation}
\beta_{FB}=0,\qquad \beta_{\pm}= \pm2\kappa \sqrt{1+\cos(\Gamma/2)\cos(k)},
\end{equation}
One band is always independent of the normalized wavenumber $k$, forming a
dispersionless FB, while the other $k$-dependent bands are
dispersive (DBs). The FB is a consequence
of diamond chain geometry. Depending on value of $\Gamma$ three different
cases are possible: one FB touching two DBs
($\Gamma=0,\, 2\pi$), one gapped FB $(\Gamma\ne 0,\, \pi,\ 2\pi)$,
three distinct FBs ($\Gamma=\pi$), the latter forming an AB cage. In a chain with $N$ unit cells each FB has an $N$-fold degeneracy in the absence of disorder and is spanned by a basis of $N$ compact, and, in general, non-orthogonal localized states (CLSs), illustrated schematically in Fig.~\ref{fig1}.

In this work we focus on the investigation of the disorder impact on the diamond chain
with fulfilled conditions for AB caging, which is characterized by
fully flat spectrum in the absence of disorder. Therefore, we are
mostly interested in dynamics of isolated compact modes and
consider two different types of initial excitations in the
network: with CLS from $\beta_{FB}=0$ and CLS from $\beta_{FB}=\pm 2$. 
We will also briefly comment in Sec.~\ref{sec:conclusion} on the behaviour of single site excitations, where a site $a$ or site $b$ from the central lattice cell are initially excited. 

To characterize the dynamics of a wavepackets and their localization,
it is useful to consider following quantities: the total intensity in each unit cell $I_n$, the second moment
$m_2$, the participation ratio $PR$, and the imbalance
$\eta$. Assuming all wavepackets are normalized such that $\sum_n I_n = 1$, these quantities are defined as
\begin{eqnarray}
I_n(z)&=&\left|a_n(z)\right|^2 + \left|b_n(z)\right|^2 + \left|c_n(z)\right|^2 ,\nonumber\\
m_1(z)& =&\sum_{n} \left( n\left|a_n(z)\right|^2+\left( n+\frac{1}{2}\right)\left(\left|b_n(z)\right|^2+\left|c_n(z)\right|^2\right) \right),\nonumber\\
m_2(z)&=&\sum_n \left(\left( n-m_1(z)\right)^2\left|a_n(z)\right|^2+ \right. \nonumber\\
& & + \left. \left(n+\frac{1}{2}-m_1(z)\right)^2\left(\left|b_n(z)\right|^2+\left|c_n(z)\right|^2\right)   \right),\nonumber\\
PR(z)&=& \left( \sum_n I_n^2 \right)^{-1}, 
\nonumber\\ 
\eta(z)&=&\sum_n{\left(\left|a_{n}(z)\right|^{2}-\left|b_{n}(z)\right|^{2}-\left|c_{n}(z)\right|^{2}\right)}, \nonumber 
\end{eqnarray}
where $\psi_n(z)=\left(a_n(z),\, b_n(z),\, c_n(z)\right)^T$ and $m_1(z)$ is the first moment or center of mass. The distribution of
$I_n$ over the lattice cells shows the efficiency of certain
disorder which is expected mostly to affect the tails of the
initially excited compact modes. The $m_2$ gives information about
the CLS or wavepacket density spreading, $PR$ about the number of
sites significantly populated by the field, while the
imbalance $\eta$ about how energy is distributed between
corresponding sub-lattices. Although the second moment and the participation ratio are linked, and broader wavepackets are expected to occupy a greater number of sites,
there are some particular situations in which this is not so. For
example in the case with self-trapping (nonlinear networks) the
second moment typically increases in time (due to unbounded spreading linear dispersive waves) but the
participation number stays more or less constant. It is related
with a frozen bulk that does not evolve with certain sites remaining highly occupied. When the noise is added to such a system the
interplay between the self-trapping and (de)localization from
disorder includes interesting features~\cite{QDSegev}.

\section{Static disorder}
\label{sec:static}

The effect of static disorder on compact modes in
the uniform diamond chain with and without an artificial magnetic field was investigated in detail in Refs.~\cite{vidal,Flach}. The behaviour depends on the degree of
isolation of the FB from DBs. In general, static disorder removes the FB
eigenvalue degeneracy and provides mixing inside this band, as
well as with states originating from any DBs. The consequences
of mixing between the FB and DB states in the
presence of weak disorder were the appearance of heavy tailed
statistics and multiple localization length scales generically
related to the existence of sparse, multi-peaked modes. In the case of gapped FB states weak disorder
was not strong enough to cause mixing of FB and DB states, so the localization length was independent of the disorder strength~\cite{nashkagome}. However, disorder affected the mode behavior
via appearance of tails in the FB states' profiles, and finally in
the presence of strong disorder the mode profiles resembled those of ordinary Anderson localization.

We consider two types of static disorder ($A=0$): 
\begin{itemize}
\item QD1: On-site disorder ($\epsilon_{jn}\ne 0, \delta \Gamma_{jn} = 0$)
\item QD2: Hopping disorder  ($\epsilon_{jn} =  0, \delta \Gamma_{jn} \ne 0$)
\end{itemize}
For weak static on-site disorder ($W<2$), the
eigenstates from each band are divided by gaps. Their localization
lengths saturate and do not depend on the disorder strength. For
strong on-site disorder,  $W>2$, mixing between energy bands
occurs, which causes decrease of the localization length with
increase of $W$. This decrease is due to the prevalence of
Anderson (exponential) localization over geometrical localization~\cite{nashkagome,PRBDesyatnikov}.

\begin{figure}
    \center\includegraphics [width=\columnwidth]{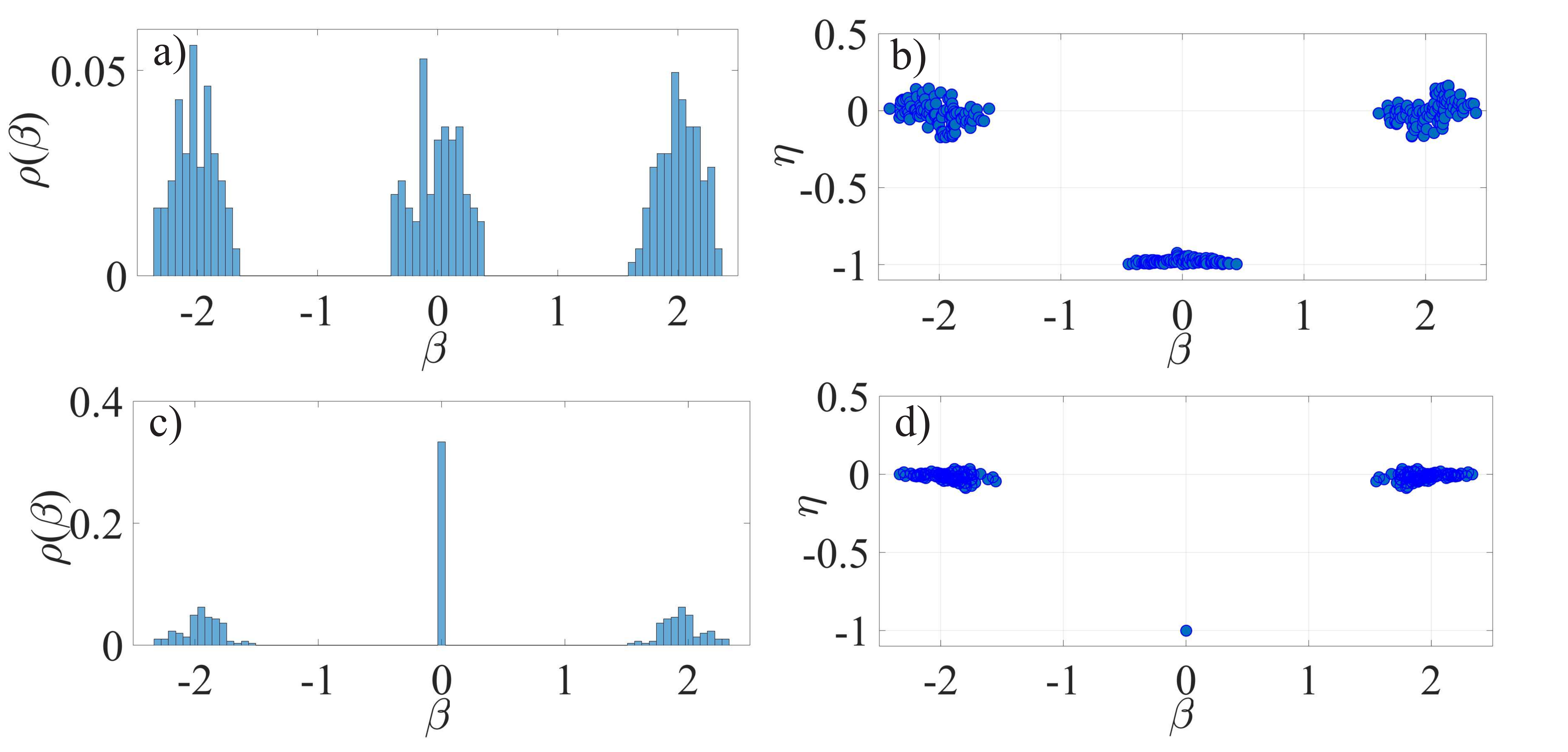} \caption{
     The eigenstate density spectrum $\rho$ (a) and corresponding imbalance $\eta$ (b) for diamond chain with $\Gamma=\pi$ (the AB
caging) in the presence of on-site static disorder of
strength $W=1$. The corresponding quantities in the presence of
static hopping disorder are plotted in (c) and (d).}
    \label{fig1a}
\end{figure}

We solve the eigenvalue problem numerically to obtain the eigenstate
spectra shown in Fig.~\ref{fig1a}, using a lattice with $N=101$ cells and disorder strength $W=1$. Without disorder the three isolated
FBs are $N$-fold degenerate. Introducing QD lifts the degeneracy. On-site disorder fully removes the degeneracy, i.e. all FBs are affected (Fig. \ref{fig1a} (a)), and the sub-lattice symmetry of the eigenstates is broken, leading to $\eta \ne 0, -1$ (Fig. \ref{fig1a} (b)). On the other hand, hopping disorder
only partially lifts degeneracy of the FBs. While the degeneracy of the two periphery FBs ($\beta_{FB}=\pm 2$) is removed, the central FB remains flat and degenerate and its eigenstates preserve their symmetry [Fig.~\ref{fig1a}(c,d)]. This is because the central FB is protected by the lattice's bipartite symmetry, which is preserved under hopping disorder.

\begin{figure}
    \center\includegraphics [width=\columnwidth]{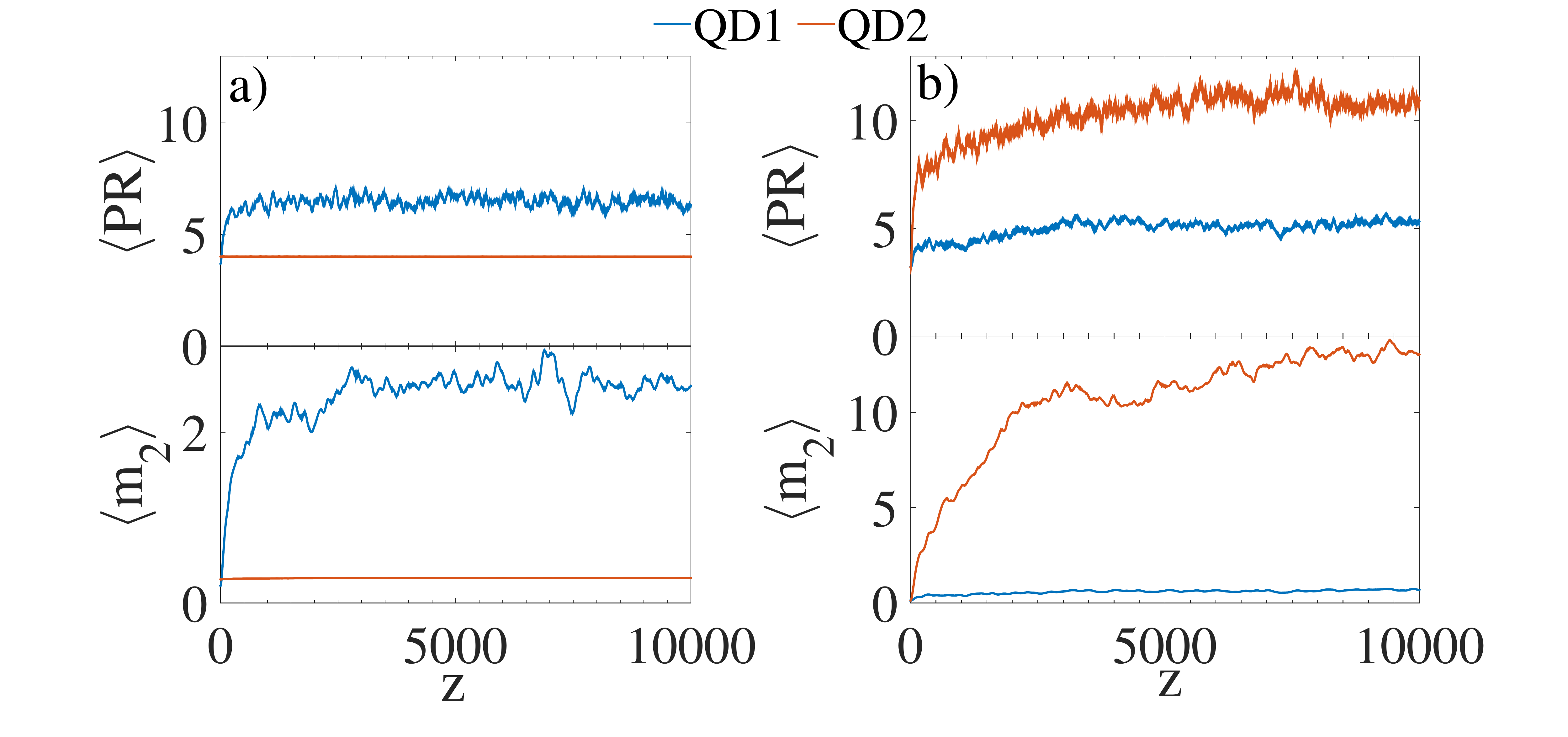}
    \caption{Dynamics of the disorder-averaged participation ratio $\langle PR \rangle$ and second moment $\langle m_{2} \rangle$ obtained for excitations of the
$\beta_{FB}=0$ (a) and $\beta_{FB} = 2$ CLS (b), obtained from $50$ realizations of static disorder.} \label{fig6}
\end{figure}

\begin{figure}
\center\includegraphics [width=\columnwidth]{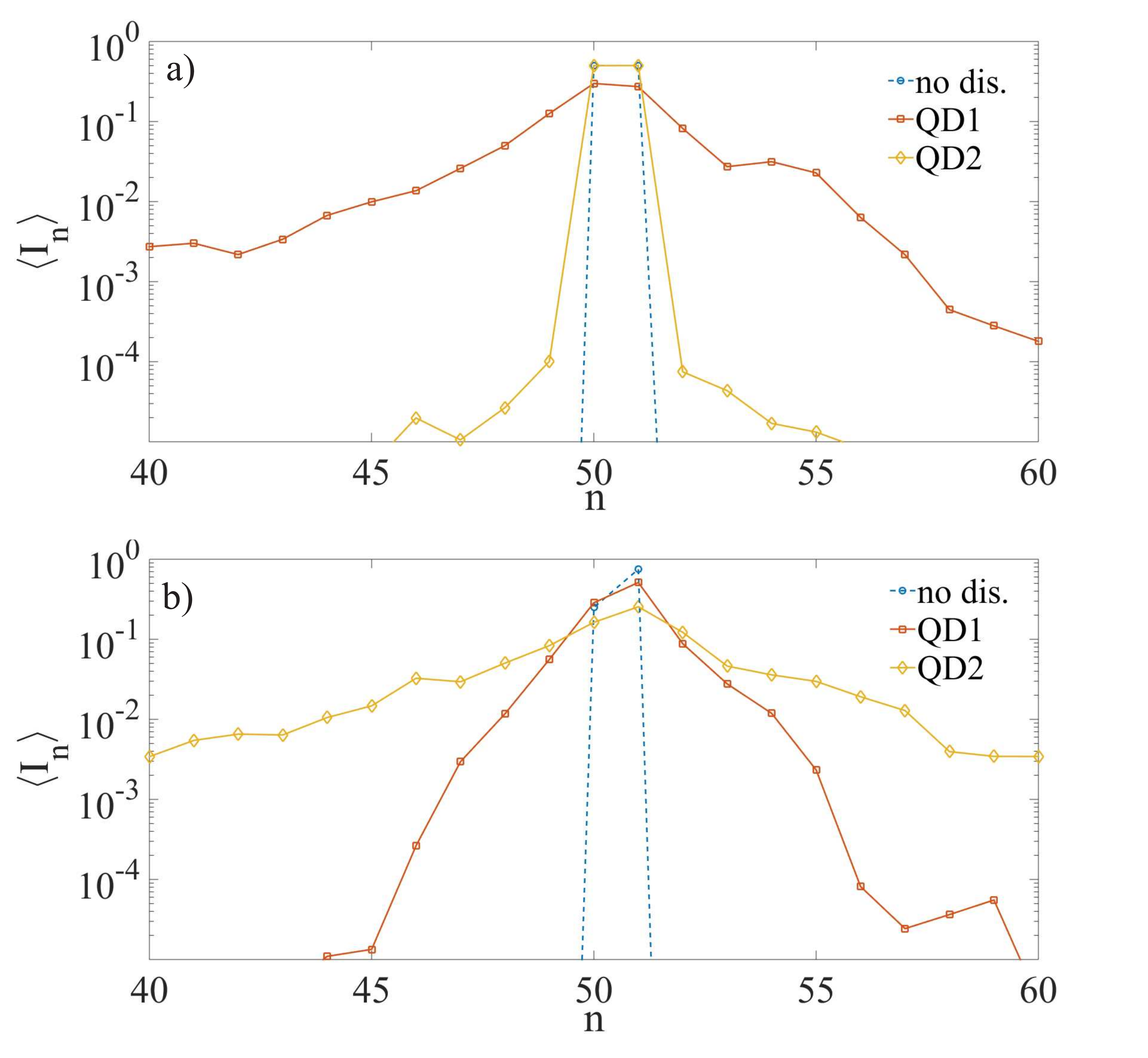}
\caption{Disorder-averaged intensity profiles $\langle I_n \rangle$ at $z=10000$ for $\beta_{FB}=0$ (a) and $\beta_{FB}=2$ (b) CLS excitations, obtained from $50$ realizations of static disorder. } \label{fig5}
\end{figure}

To reveal the impact of the disorder on the propagation dynamics in the AB cage we numerically solve Eq.~\eqref{system6}, taking the normalized compact modes of Fig.~\ref{fig1} as initial conditions. We fix the disorder strength $W=0.1$ and consider an ensemble of $50$ disorder realizations. Fig.~\ref{fig6} illustrates the dynamics of the disorder-averaged $\langle PR \rangle$ and $\langle m_2 \rangle$, which saturate to finite values indicating the onset of Anderson localization. Fig.~\ref{fig5} additionally plots the ensemble-averaged beam intensity profile $\langle I_n \rangle$ after a propagation distance of $z=10000$ a.u. Consistent with the numerically-calculated eigenvalue spectra, we observe that the compact localization present in the ideal (non-disordered) system is destroyed, replaced by exponential Anderson localization. Moreover, the $\beta_{FB} = 0$ CLS is much less sensitive to the hopping disorder, which does not spread at all and preserves its initial profile.

\begin{figure}
    \center\includegraphics [width=\columnwidth]{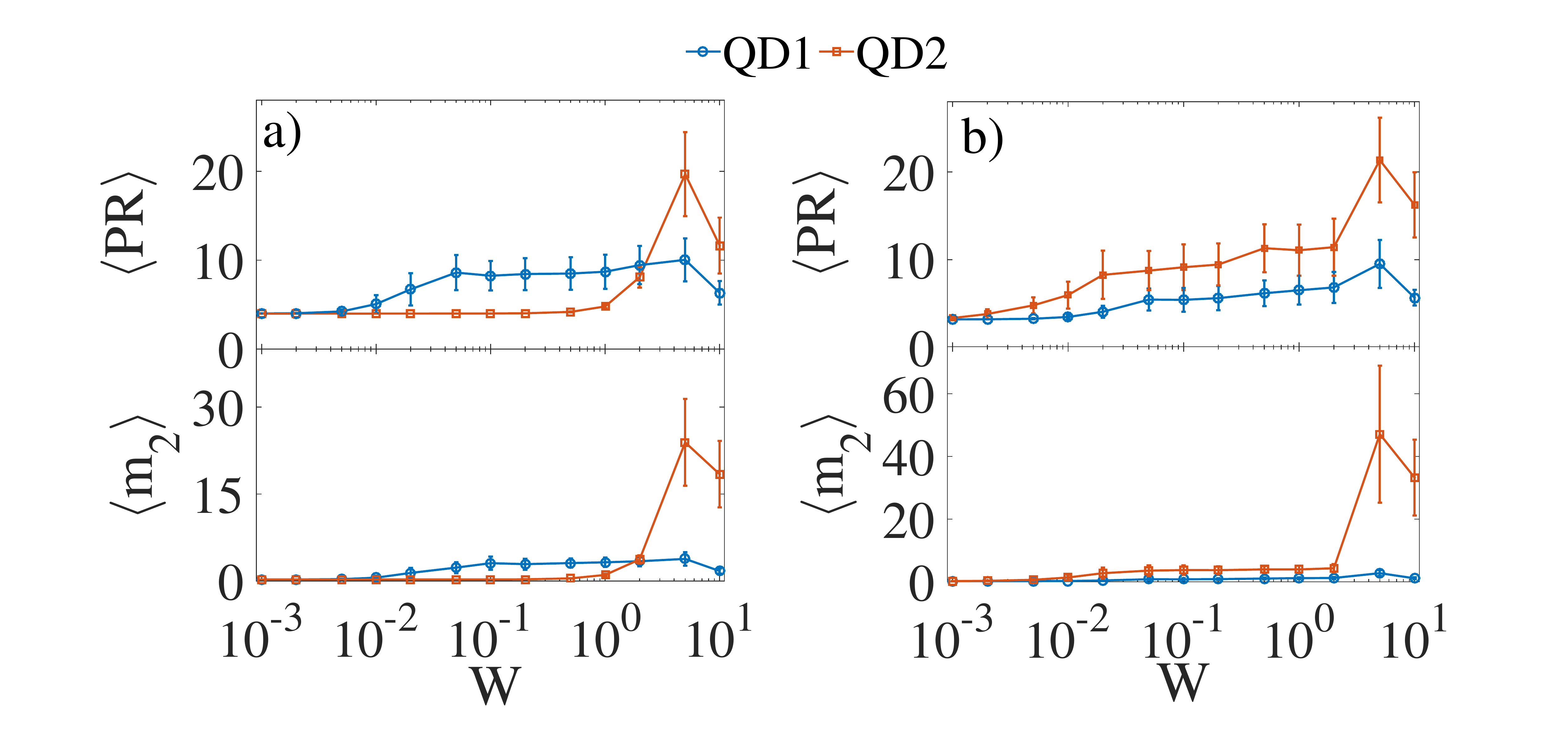} \caption{
	Disorder-averaged $\langle PR \rangle$ (upper plot) and $\langle m_2 \rangle$ (lower plot)
	after propagation length $z=1000$ for different stationary QD realizations
	vs. disorder strength $W$. (a) CLS from central FB ($\beta_{FB}=0$) and (b) CLS
	from the periphery FB are initially excited in the lattice (from $\beta_{FB}=2$ ). 	Error bars illustrate the standard deviations of corresponding quantities.
 }
    \label{fig0}
\end{figure}

Fig.~\ref{fig0} quantifies the dependence of saturated disorder averages $\langle PR \rangle$ and $\langle m_2 \rangle$ on the disorder strength $W$. For very weak disorder ($W<0.01$) $\langle PR \rangle$ and $ \langle m_2 \rangle$ of initially injected CLS from $\beta_{FB}=0$ are saturated to the corresponding values in the absence of
disorder. This is the case for both on-site and hopping disorder. While this
tendency continues for the hopping disorder for stronger $W$, the on-site disorder starts to
affect the CLS from $\beta_{FB}=0$, which is slightly smeared over the neighbouring
cells, corresponding to saturation of $\left\langle PR\right\rangle $ and $\left\langle m_2\right\rangle $ at higher values. For strong disorder $W \gtrsim 2$ we observe an increase in the spreading due to a transition from AB cage-induced localization to regular Anderson localization. 

\section{Periodic disorder}
\label{sec:periodic}
For periodically-evolving disorder an important new energy scale emerges: the ratio of the characteristic period of disorder to the width of the band gap in the Bloch wave spectrum. We start by considering three classes of on-site periodic disorder (taking $A=1$ and $\delta \Gamma = 0$), 
\begin{itemize}
\item QDP1: driving resonant with the band gap ($\omega_0 = 2$)
\item QDP2: off-resonant driving ($\omega_0 = 0.1$)
\item QDP3: driving resonant with twice the band gap ($\omega_0 = 4$)
\end{itemize}
The resonant driving cases QDP1 and QDP3 can induce strong mixing between different elementary CLSs to create new Floquet eigenstates. We are interested in how this disordered mixing affects the wavepacket localization.

\begin{figure}
	\center\includegraphics [width=\columnwidth]{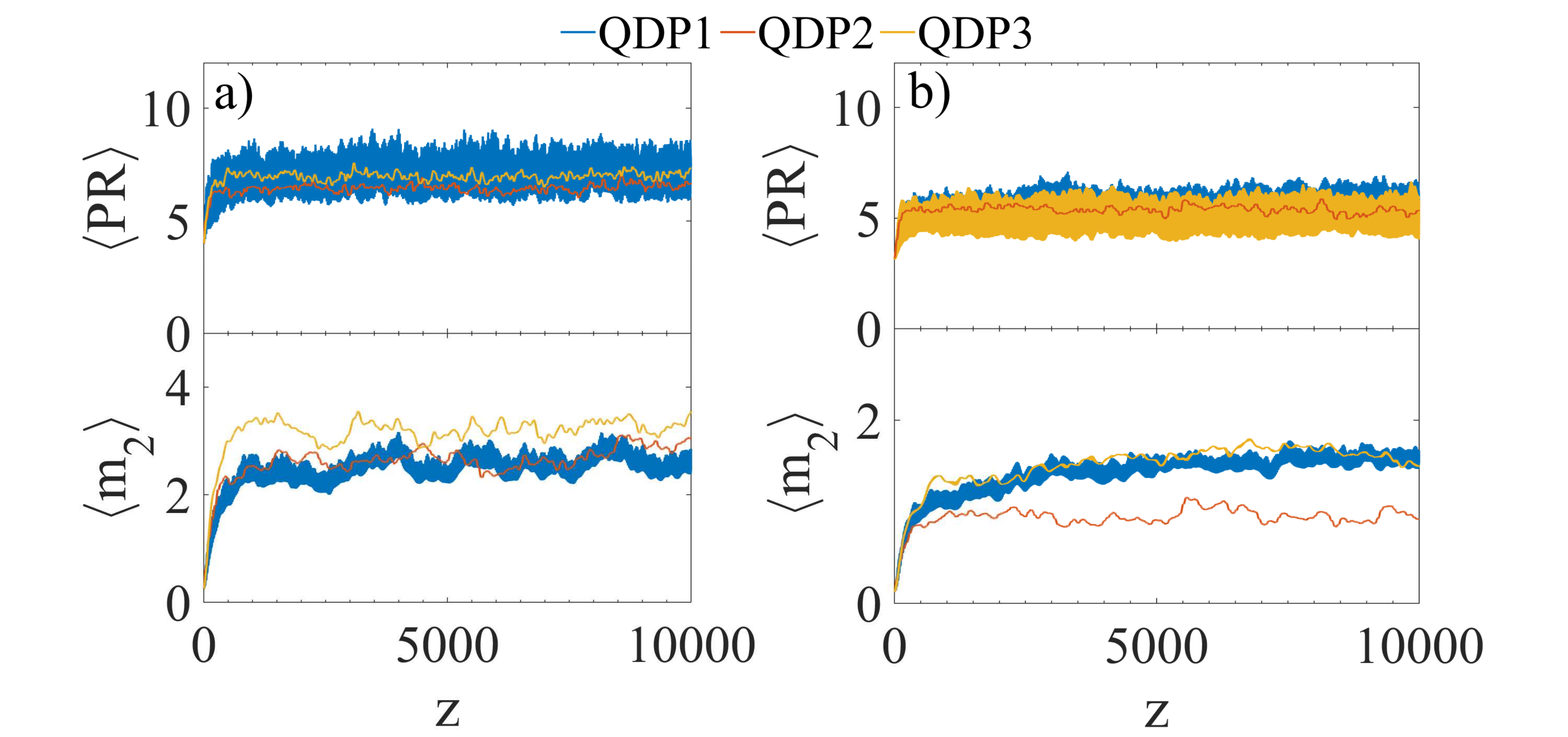}
	 \caption{
Dynamics of the disorder-averaged participation ratio $\langle PR \rangle$ and second moment $\langle m_2 \rangle$ obtained for excitations of the
$\beta_{FB}=0$ (a) and $\beta_{FB} = 2$ CLS (b), averaged over $50$ realizations of the three classes of periodic disorder.}

\label{qdp1}
\end{figure}

\begin{figure}
	\center\includegraphics [width=\columnwidth]{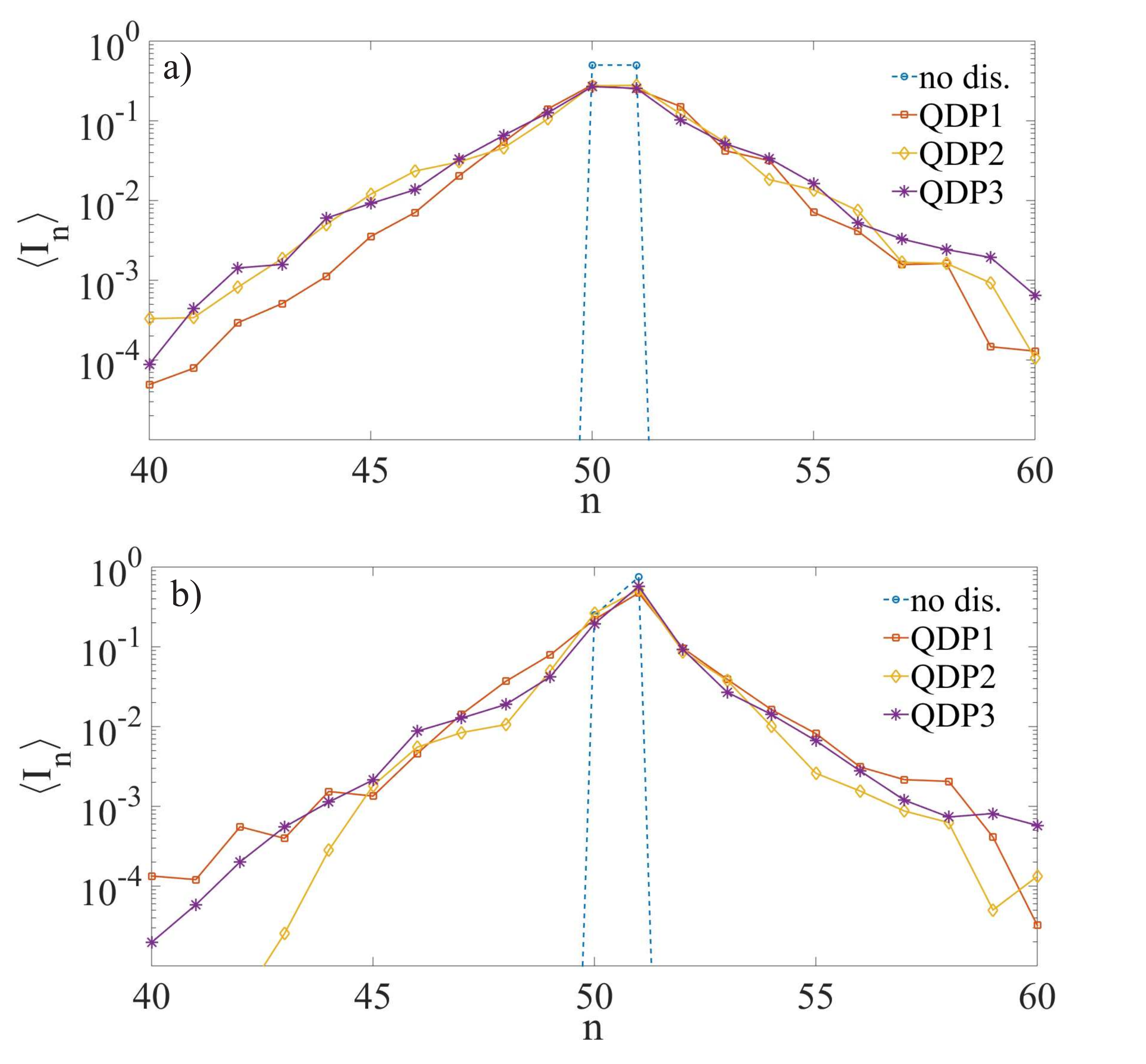} 
		\caption{
Intensity profiles $\langle I_n \rangle$ at $z=10000$ for $\beta_{FB}=0$ (a) and $\beta_{FB}=2$ (b) CLS excitations, averaged over $50$ realizations of the three classes of periodic disorder.}
\label{qdp2}
\end{figure}

Fig.~\ref{qdp1} shows the long time dynamics of the disorder-averaged participation number $\langle PR \rangle$ and the second moment $\langle m_2 \rangle$. Both $\langle PR \rangle$ and $\langle m_2 \rangle$ saturate after an initial transient spreading, regardless of the type of periodic disorder or the initial CLS. Localization after that persists. Under resonant driving additional rapid oscillations in $\langle PR \rangle$ and $\langle m_2 \rangle$ appear due to interband coupling. This occurs for QDP1 when the central FB is initially excited, and for both QDP1 and QDP3 when one of the peripheral FBs is excited. The averaged intensity profiles in Fig.~\ref{qdp2} show similar exponential tails in all cases.

\begin{figure}
	\center\includegraphics [width=\columnwidth]{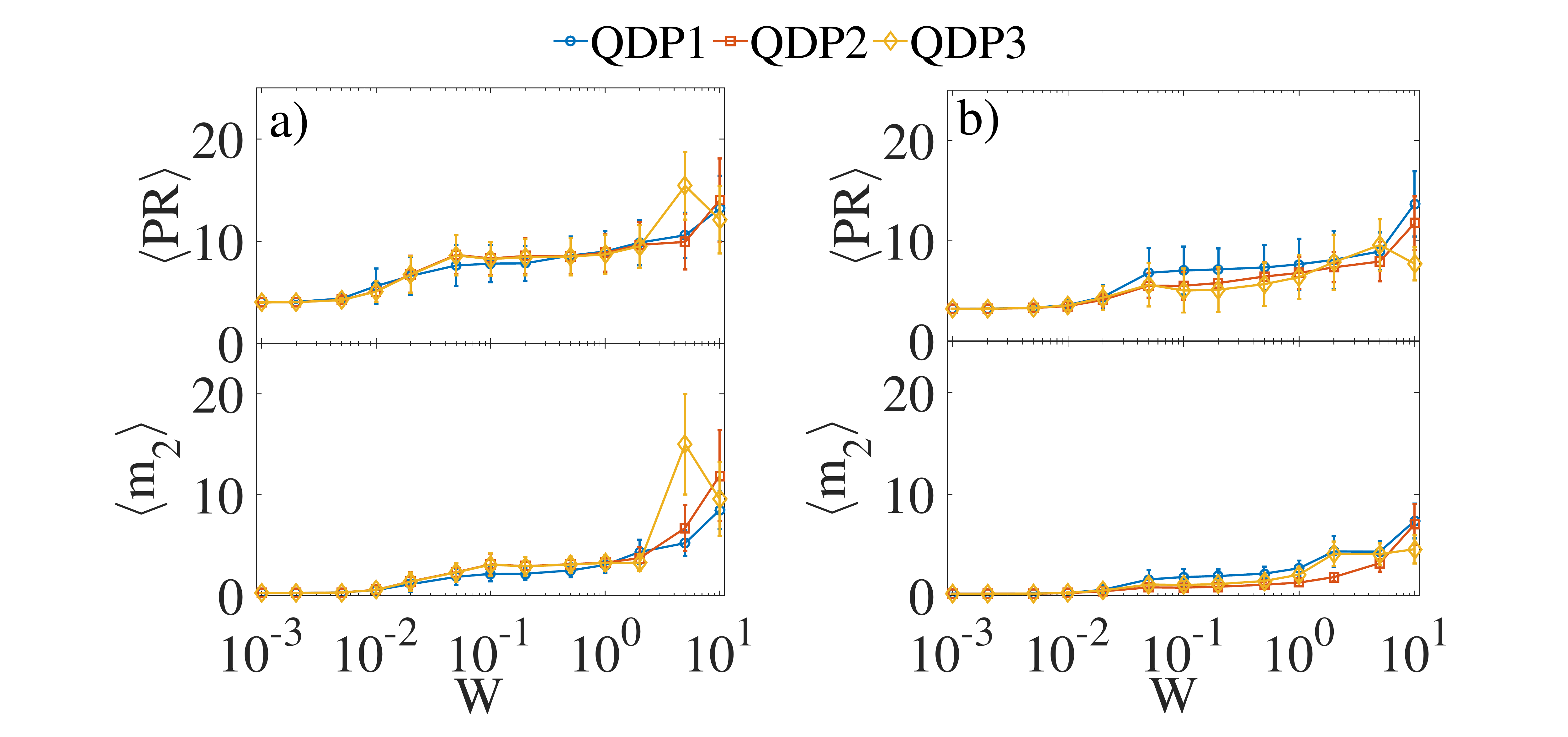} \caption{
	The averaged value of $\langle PR \rangle$ (upper plot) and $\langle m_2 \rangle$ (lower plot)
	over the propagation length $z=1000$ for different stationary QD realizations
	vs. disorder strength $W$. (a) CLS from central FB ($\beta_{FB}=0$) and (b) CLS
	from the periphery FB ($\beta_{FB}=2$) are initially excited in the lattice. }
\label{qdp3}

\end{figure}

We can understand the effect of the periodic disorder on the wavepacket spreading following the approach in Refs.~\cite{teza,nashkagome}. Based on perturbation theory for linear systems in the presence of time-dependent perturbations, we expect to see transitions between different static eigenstates. The strength of these transitions is determined by the spatial overlap between the states, and whether the frequency of the perturbation is resonant with the energy difference between the states. Due to the strong localization induced by the AB caging, there can only be appreciable overlap with directly neighboring CLS, which have random energies due to the static part of the disorder. Consequently, while resonances between neighboring states can in principle lead to slight expansion of the wavepacket, this occurs with low probability. Therefore the wavepackets remain bounded and strongly localized. Described findings are clearly seen in Fig. (\ref{qdp3}) where the effect of periodic QDs of different strength on the CLS spreading is presented. In general slowly increasing slopes of the curves $\left\langle PR\right\rangle $ vs. $W$ and $\left\langle m_2\right\rangle $ vs. $W$  are consequences of the time dynamics caused by the periodic modulation of on-site disorder. 

\begin{figure}
	\center\includegraphics [width=\columnwidth]{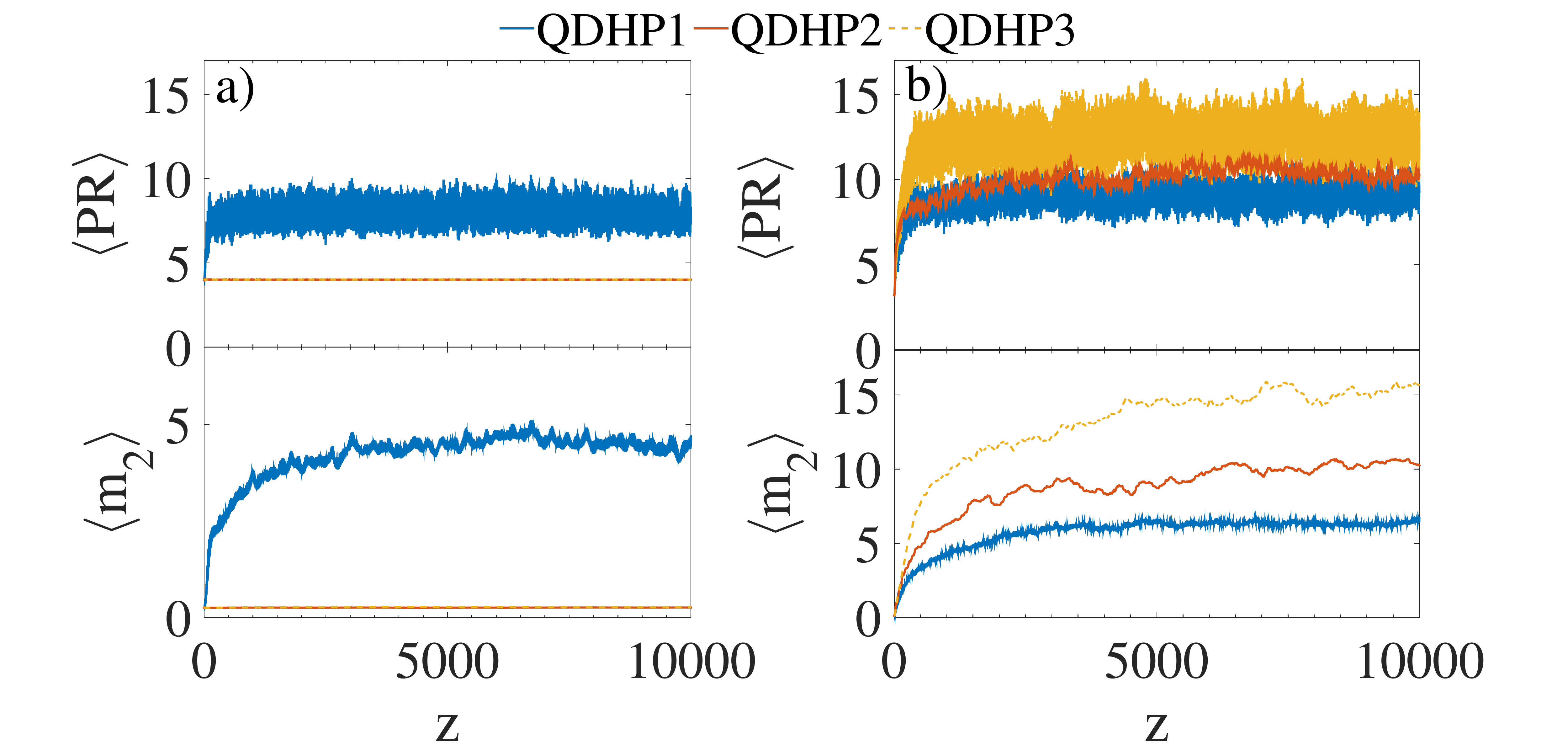}
	\caption{
		Dynamics of the disorder-averaged participation ratio $\langle PR \rangle$ and second moment $\langle m_2 \rangle$ obtained for excitations of the
		$\beta_{FB}=0$ (a) and $\beta_{FB} = 2$ CLS (b), averaged over $50$ realizations of the three classes of periodically modulated hopping disorder.}
	
	\label{hqdp1}
\end{figure}

\begin{figure}
	\center\includegraphics [width=\columnwidth]{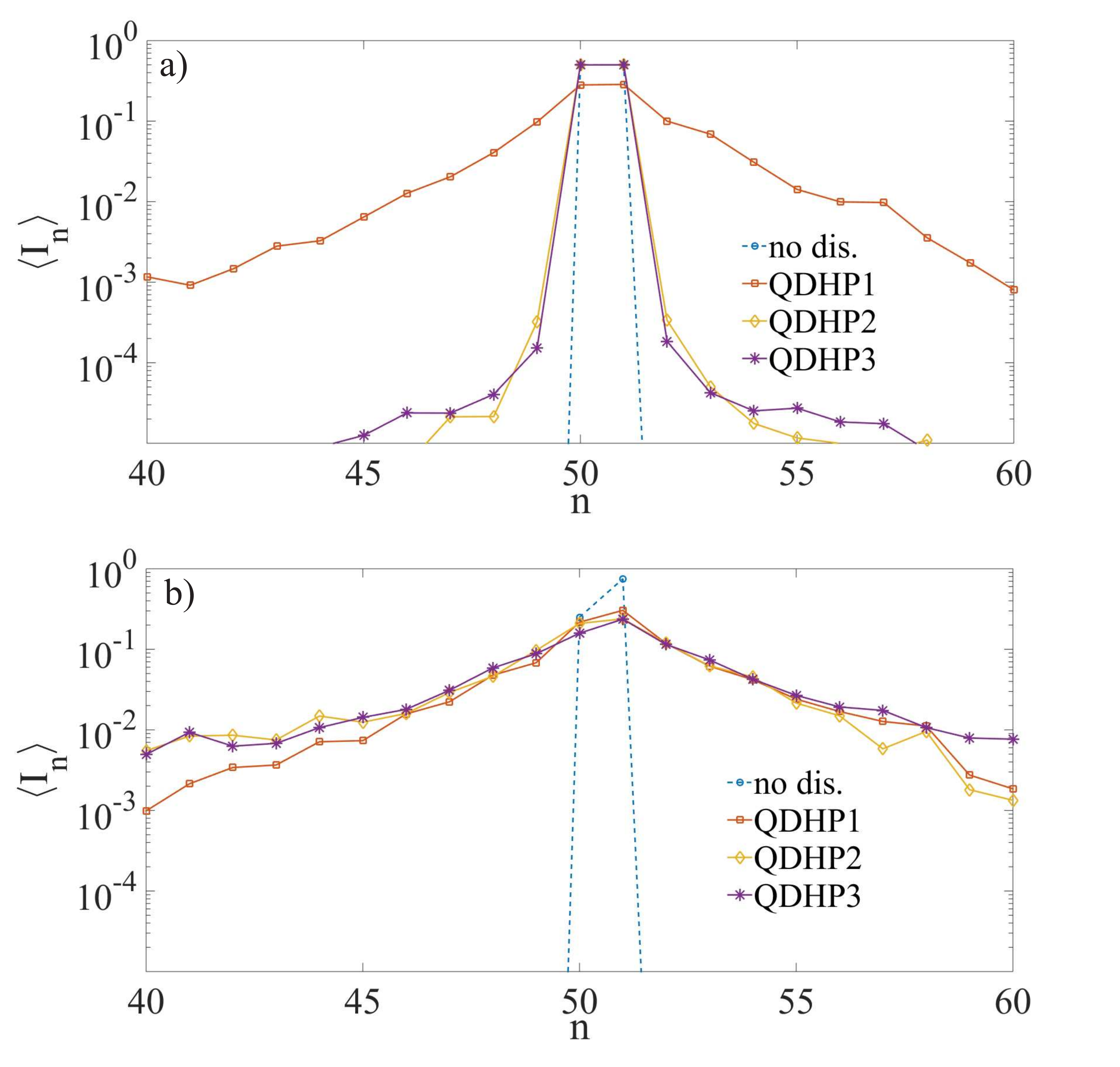} 
	\caption{
		Intensity profiles $\langle I_n \rangle$ at $z=10000$ for $\beta_{FB}=0$ (a) and $\beta_{FB}=2$ (b) CLS excitations, averaged over $50$ realizations of the three classes of periodically modulated hopping disorder.}
	\label{hqdp2}
\end{figure}

Next we consider the time-periodic hopping disorder ($A=1$):
\begin{itemize}
\item QDHP1: driving resonant with the band gap ($\omega_0 = 2$)
\item QDHP2: off-resonant driving ($\omega_0 = 0.1$)
\item QDHP3: driving resonant with twice the band gap ($\omega_0 = 4$)
\end{itemize}
The corresponding results are shown in Figs. (\ref{hqdp1}) and (\ref{hqdp2}), which illustrate the evolution of the disorder averaged $PR$, $m_2$, and intensity profiles $\langle I_n \rangle$ for both types of the compact localized modes, respectively. The particularity of the case with time periodic hopping disorder is clear difference between the effect of resonant with the band gap ($\omega_0 = 2$) driving and other types of drivings of the CLS from $\beta_{FB}=0$ band. While the resonant driving (QDHP1) induces exponential tails in the localized mode, the off-resonant ones (QDHP2 and QDHP3) preserve the strong localization of the $\beta_{FB} = 0$ CLS, which remains almost compact (plots (a) in Figs. (\ref{hqdp1}) and (\ref{hqdp2})). This is straightforward consequence of the wavepackets relaxation via the bottleneck path through the $a$ sublattice in the resonant QDHP1 case. Regarding the CLS initiated from the $\beta_{FB}=2$, all types of modulated hopping disorder induced the fast growth and saturation of $\left\langle PR\right\rangle $ and $\left\langle m_2\right\rangle $ to the finite values (Fig. (\ref{hqdp1}) (b)) and the exponential localization (Fig. (\ref{hqdp2}) (b)). Again the observed dynamical properties can be associated with the active role of the $a$ sublattice, which is in this case generically whole time 'populated' and thus 'introduced' in the mode relaxation.  
 
\section{Non-quenched disorder}
\label{sec:nqd}

\begin{figure}[h]
    \includegraphics [width=\columnwidth]{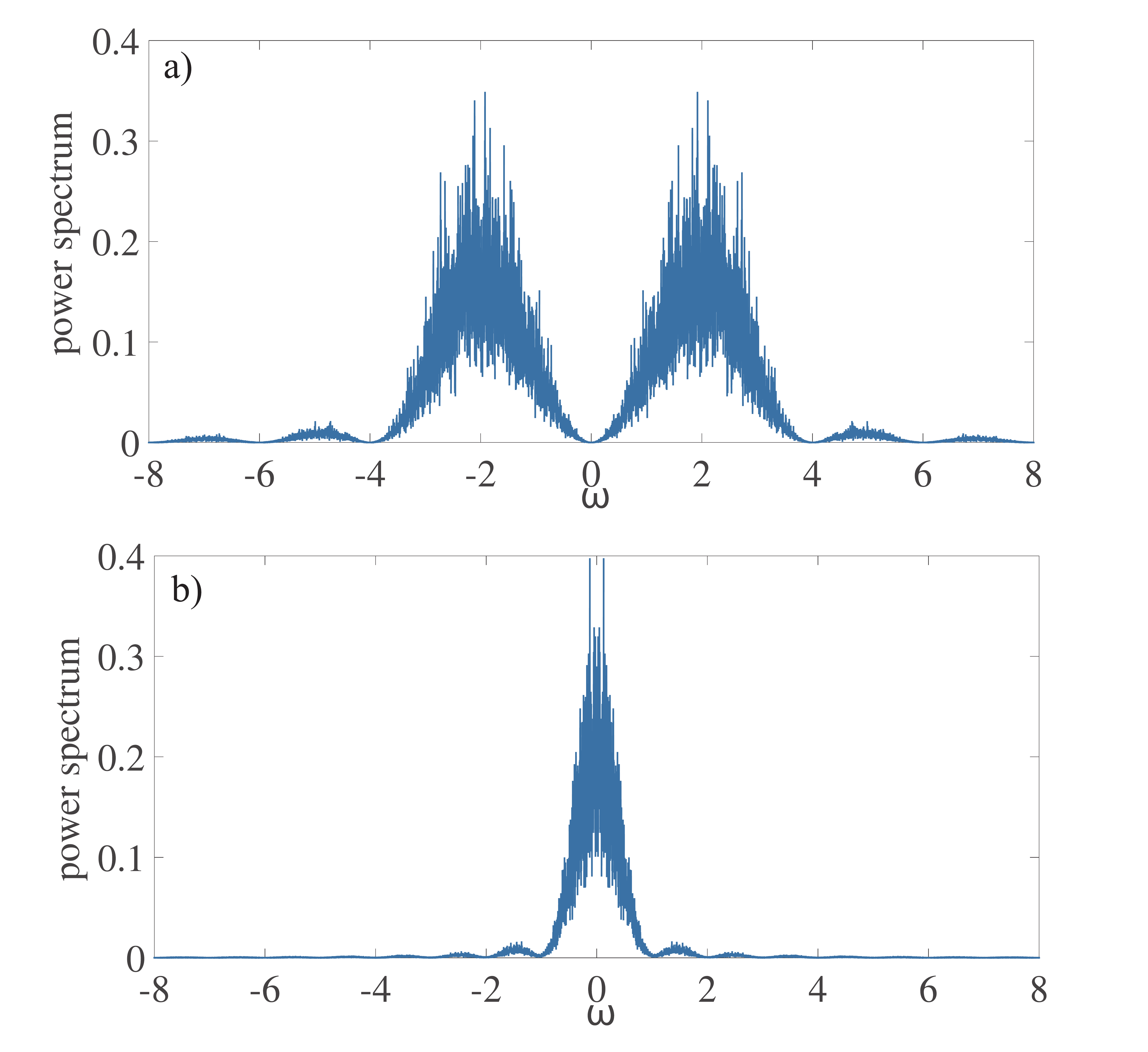} \caption{The on-site NQD
        power spectra obtained after averaging over $10$ different realizations of the set of random numbers for
        (a) NQD1 and (b) NQD2 types.} \label{nqdpr}
\end{figure}

Finally, we study the NQD, which introduces an additional characteristic scale, the dephasing length $\Delta z$. The monochromatic perturbations studied in the previous section are now broadened to have a finite bandwidth, enabling coupling between successive CLS and wavepacket spreading. The efficiency of the spreading is dictated by the NQD power spectrum, shown in Fig.~\ref{nqdpr} for onsite disorder (with amplitude $A=1$). We consider the following types of NQD disorder:
\begin{itemize}
\item NQD1: On-site disorder $\omega_0 = 2, \Delta z = 5$
\item NQD2: On-site disorder $\omega_0 = 0.1, \Delta z = 10$
\item NQD3: Hopping disorder $\omega_0 = 0, \Delta z = 10$.
\end{itemize}

\begin{figure}[h]
    \center\includegraphics [width=\columnwidth]{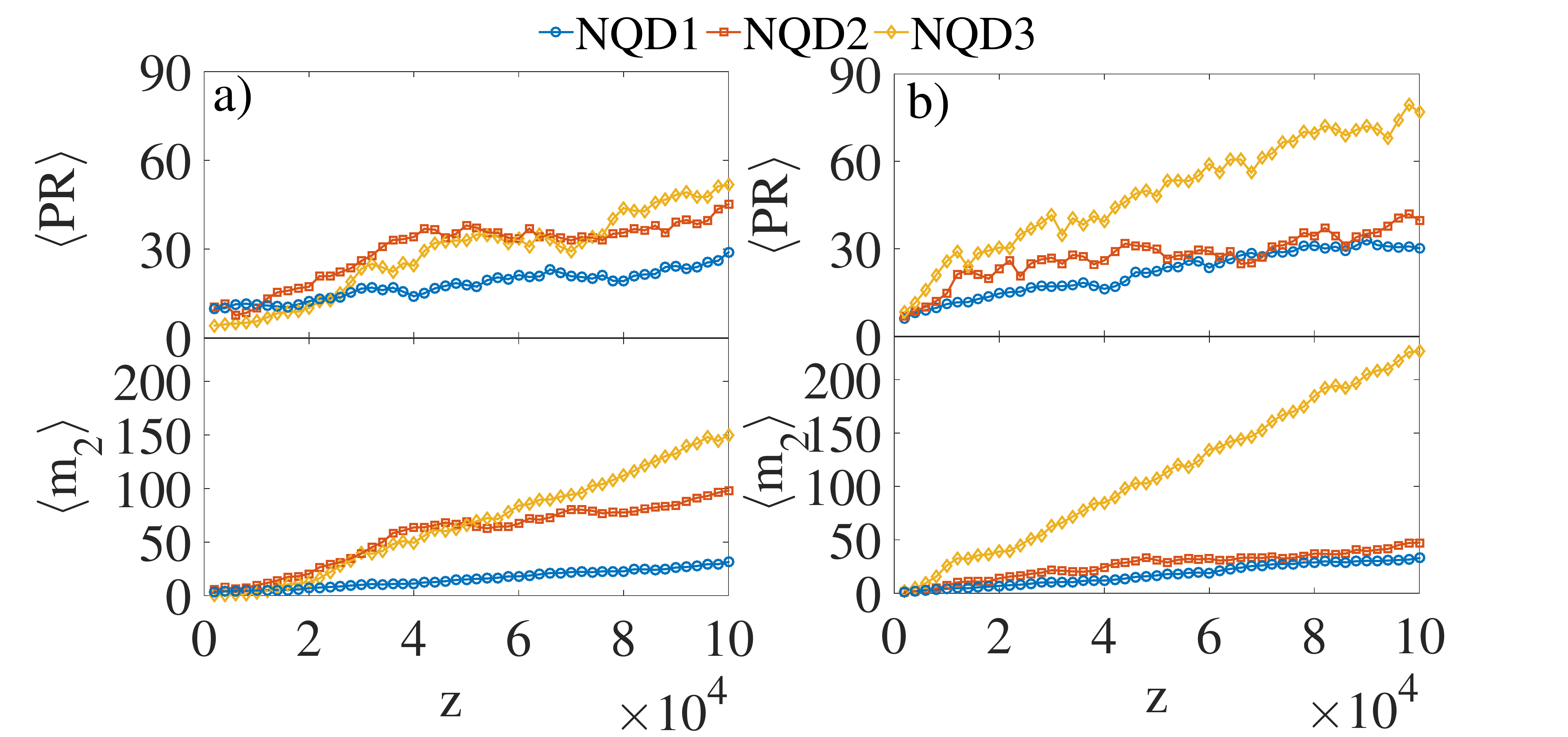}
\caption{Averaged values of participation ratio $\langle PR \rangle$ and second moment $\langle m_{2} \rangle$
    over $50$ propagation windows, which are obtained for
    CLS from $\beta_{FB}=0$ excitation (a) and the CLS from $\beta_{FB}=2$
    excitation (b) and different types of NQDs. The mentioned
    propagation windows are obtained by dividing the total propagation
    length into $50$ parts of equal length.} \label{fig8}
\end{figure}

The NQD type of disorder is characterized
by abrupt changes of the disorder realization along the
propagation directions which occur at regular distances.
These ``kicks'' have impacts on the spreading of the wavepacket,
which reflect on the averaged participation number $\langle PR \rangle$ and the second
moment $\langle m_2 \rangle$ behaviour (Figs. \ref{fig8}). Both
quantities grow with $z$, regardless of the type
of NQD or CLS excitation. This indicates that NQD breaks localization
and causes wave-packet spreading in all cases. Comparing $\langle PR \rangle$ and $\langle m_2 \rangle$, we
find faster spreading for hopping disorder (NQD3) than in the cases with on-site disorder (NQD1,NQD2). This difference is stronger for CLS from $\beta_{FB}=2$  than CLS from $\beta_{FB}=0$ excitation. Additionally, after an initial transient the rate of spreading measured by $\langle m_2 \rangle$ for NQD1 and NQD2 tends to slow, but not
in the NQD3 case. Furthermore, for the CLS from $\beta_{FB}=0$  excitation faster spreading occurs for NQD2 compared to NQD1, which we attribute to the stronger overlap of the disorder spectrum with the static modes.

\section{Conclusion}
\label{sec:conclusion}

The purpose of this study was to reveal the effect of different
types of quenched and non-quenched disorders on the dynamics of
compact localized excitations (CLSs) in the diamond chain threaded by an effective magnetic flux, which forms an Aharonov-Bohm cage with a completely flat spectrum. These classes of disorder are particularly relevant to recent experimental realizations of AB cages in photonic waveguide arrays~\cite{Mukherjee,Szameit}. We have found that the CLS become exponentially localized under quenched and periodic disorders. Notably, strong localization under periodic disorder persists even if the periodic modulation resonantly couples different bands. Abrupt changes of the disorder realization forming non-quenched disorder destroys localization and induces wavepacket spreading for both on-site and hopping disorders, with the latter resulting in more rapid spreading. 

We focused on the dynamics of the CLS hosted by the flat bands of the diamond chain. Such CLS are obtained via excitation of multiple waveguides. We found qualitatively similar localization behaviour for single waveguide excitations, since they can be expressed as superpositions of the CLS. Namely, an $a$ waveguide excitation is a superposition of CLS from the $\beta_{FB} = \pm 2$ bands, while a $b$ waveguide excitation involves CLS from all the FBs. In both cases the localization or spreading dynamics are accompanied by rapid oscillations due to interference between the different bands. 

In the future, it would be interesting to generalize this study to two- and three-dimensional Aharonov-Bohm cage lattices~\cite{vidal}, where the dynamics are expected to be more sensitive to the disorder strength. We considered idealised tight binding models of quenched and non-quenched disorder which we believe are representative of a variety of disordered systems including waveguide arrays and optical lattices for cold atoms. To determine the conditions under which the delocalization or saturation of wavepacket spreading may be observable in experiment, it will be necessary to conduct more rigorous simulations taking into account the rapid periodic modulation~\cite{Mukherjee} or auxiliary sites~\cite{Szameit} used to create the synthetic magnetic field.

\begin{acknowledgments}  We acknowledge support from the
	Ministry of Education, Science and Technological Development of
	the Republic of Serbia (Project No. III 45010) and the Institute for Basic Science in Korea (IBS-R024-Y1).
\end{acknowledgments}


\begin{thebibliography}{99}
	
		\bibitem{dd} R. S. MacKay, L. Vazquez, and M. P. Zorzano, {\it Localization and Energy Transfer in Nonlinear Systems}, (World Scientific Publishing Co. Pte. Ltd., Singapore, 2003).
	
	\bibitem{dd1} D. K. Cambell, S. Flach, and Yu. S. Kivshar, Localizing energy through nonlinearity and discretness, Physics Today \textbf{57}, Issue 1, 43 (2004).

        \bibitem{Anderson1958} P. W. Anderson, Absence of diffusion in certain random lattices, Phys. Rev. \textbf{109}, 1492 (1958).
        
            
        
        
        \bibitem{SegevHypertransport} L. Levi, Y. Krivolapov, S. Fishman, and M. Segev, Hypertransport
        of light and stochastic acceleration by evolving
        disorder, Nat. Physics \textbf{8}, 912 (2012).
        
        \bibitem{SegevSuperdiffusion} Y. Krivolapov, L. Levi, S. Fishman, M.
        Segev, and M. Wilkinson, Super-diffusion in optical realizations of Anderson localization,  New J. Phys. \textbf{14}, 043047 (2012).
        
        
        \bibitem{p21} O. V. Kibis, H. Sigurdsson, and I. A.
        Shelykh,  Aharonov-Bohm effect for excitons in a semiconductor
        	quantum ring dressed by circularly polarized light, Phys. Rev. B
        \textbf{91}, 235308 (2015).
        
        \bibitem{quantrings} M. Hasan, I. V. Iorsh, O. V. Kibis, and I. A.
        Shelykh, Optically controlled periodical chain of quantum
        	rings, Phys. Rev. B \textbf{93}, 125401 (2016).
        
        
        
        
        \bibitem{moller}
        G. M\"oller and N. R. Cooper, Synthetic gauge fields for lattices with multi-orbital unit cells: routes towards a $\pi$-flux dice lattice with flat bands, New J. Phys. {\bf 20}, 073025 (2018).
        
        
        \bibitem{fang}
        K. Fang, Z. Yu, and S. Fan, Realizing effective magnetic field for photons by controlling the phase of dynamic modulation, Nature Photon. {\bf 6}, 782 (2012).
        
        

        \bibitem{FB_review} 
        D. Leykam, A. Andreanov, and S. Flach, Artificial flat band systems: from lattice models to experiments, Adv. in Phys.:X \textbf{3}, 1470352 (2018).
        
        \bibitem{FB_review2} D. Leykam, and S. Flach, Perspective: Photonic flatbands, Photon. {\bf 3}, 070901 (2018).
   
   
   
        

        
        \bibitem{FlachNL} S. Flach, D. O. Krimer, and C. Skokos, Universal Spreading
        of Wave Packets in Disordered Nonlinear Systems, Phys. Rev. Lett. {\bf 102}, 024101 (2009).
        
        \bibitem{nasabc} G. Gligori\'{c}, P. P. Beli\v{c}ev, D. Leykam, and A. Maluckov, Nonlinear symmetry breaking of Aharonov-Bohm cages, Phys. Rev. A \textbf{99}, 013826 (2019).  
  
        
		\bibitem{diliberto} M. Di Liberto, S. Mukherjee, N. Goldman, arXiv:1810.07641.        
          
  
        
        \bibitem{EPLFlach} S. Flach, D. Leykam, J. D. Bodyfelt, P.
        Matthies, and A. S. Desyatnikov, Detangling flat bands into Fano lattices, Europhys. Lett. \textbf{105},
        30001 (2014).
        
        \bibitem{Flach} D. Leykam, J. D. Bodyfelt, A. S. Desyatnikov, and
        S. Flach, Localization of weakly disordered flat band states, Eur. Phys. J. B \textbf{90}, 1 (2017).
        
        \bibitem{Gneiting2018} C. Gneiting, Z. Li, and F. Nori, Lifetime of flatband states, Phys. Rev. B {\bf 98}, 134203 (2018).        
               
               	\bibitem{abohm} Y. Aharonov and D. Bohm, Significance of electromagnetic potentials in the quantum theory, Phys. Rev. \textbf{115} 485 (1959).
               
         \bibitem{vidal} J. Vidal, R. Mosseri, and B. Dou\c{c}ot, Aharonov-Bohm cages in two-dimensional structures, Phys. Rev. Let. \textbf{81}, 5888 (1998).
        
        \bibitem{vidal2}  C. C. Abilio, P. Butaud, Th. Fournier, B. Pannetier, J. Vidal, S. Tedesco, and B. Dalzotto, Magnetic field induced localization in a two-dimensional superconducting wire network, Phys. Rev. Lett. \textbf{83}, 5102
        (1999).
		
        \bibitem{ahr} S. Longhi, Aharonov-Bohm photonic cages in waveguide and coupled resonator lattices by synthetic magnetic fields, Opt. Lett. \textbf{39}, 5892
        (2014).

        \bibitem{modulatedPL} I. L. Garanovich, S. Longhi, A. A. Sukhorukov, and Yu. S. Kivshar, Light propagation and localization in modulated photonic lattices and waveguides, Phys. Rep. \textbf{518}, 1 (2012).

        \bibitem{QDSzamait} L. Martin, G. Di Giuseppe, A. Perez-Leija, R. Keil, F. Dreisow, M. Heinrich, S. Nolte, A. Szameit, A. F. Abouraddy, D. N. Christodoulides, and B. E. A. Saleh, Anderson localization in optical waveguide arrays with off-diagonal coupling disorder, Opt. Express \textbf{19}, 13636 (2011).
        
         \bibitem{Longinonherm} S. Longhi, D. Gatti, G. D. Valle, Non-Hermitian transparency and one-way transport in low-dimensional lattices by an imaginary gauge field, Phys. Rev. B 92, 094204 (2015).
        
         \bibitem{Mukherjee} S. Mukherjee, M. Di Liberto, P. \"Ohberg, R. R.
        Thomson,  and N. Goldman, Experimental observation of Aharonov-Bohm cages in photonic lattices, Phys. Rev. Lett. {\bf 121}, 075502 (2018). 
	
	    \bibitem{Szameit}M. Kremer, I. Petrides, E. Meyer, M. Heinrich, O. Zilberberg, and A. Szameit, Non-quantized square-root topological insulators: a realization in photonic Aharonov-Bohm cages, arXiv:1805.05209.
	        
        
        \bibitem{26}
        C. Naud, G. Faini, and D. Mailly, Aharonov-Bohm cages in 2D normal metal networks, Phys. Rev. Lett. \textbf{86}, 5104 (2001).
        
        \bibitem{27} I. M. Pop, K. Hasselbach, O. Buisson, W. Guichard, B.
        Pannetier, and I. Protopopov, Measurement of the current-phase relation in Josephson junction rhombi chains, Phys. Rev. B \textbf{78}, 104504 (2008).



	    \bibitem{nashkagome} A. Radosavljevi\'{c}, G. Gligori\'{c}, P. P. Beli\v{c}ev, A. Maluckov, and M. Stepi\'{c}, Light propagation in binary kagome ribbons with evolving disorder, Phys. Rev. E \textbf{96}, 012225 (2017).
	
		\bibitem{teza} M. Moratti, Transport phenomena in disordered
		time-dependent potentials, Doctoral dissertation, Universit\`{a}
		degli Studi di Firenze, 2014.
		
        \bibitem{QDSegev} T. Schwartz, G. Bartal, S. Fishman, and M. Segev, Transport and Anderson localization in disordered two-dimensional photonic lattices, Nature \textbf{446}, 52 (2007).	
        
        \bibitem{PRBDesyatnikov} D. Leykam, S. Flach, O. Bahat-Treidel, and A.S.
        Desyatnikov, Flat band states: Disorder and nonlinearity, Phys. Rev. B \textbf{88}, 224203 (2013).	
        

\end{thebibliography}
\end{document}